\documentclass{spie}
\usepackage{graphicx}
\usepackage{subfigure}
\usepackage{color}
\usepackage{amsmath, amsfonts}

\newcommand{\mycitet}[1]{[\citenum{#1}]}
\newcommand{\mycitep}[1]{\cite{#1}}

\newcommand{\myfig}[1]{Fig. \ref{#1}}

\title{Electric Field Conjugation with the Project 1640 coronagraph}
\author{Eric Cady\supit{a}\footnote{\:\, eric.j.cady@jpl.nasa.gov}, 
Christoph Baranec\supit{b},
Charles Beichman\supit{abc},
Douglas Brenner\supit{d},
Rick Burruss\supit{a},
Justin Crepp\supit{e},
Richard Dekany\supit{b},
David Hale\supit{b},
Lynne Hillenbrand\supit{b},
Sasha Hinkley\supit{b},
E. Robert Ligon\supit{a},
Thomas Lockhart\supit{a},
Ben R. Oppenheimer\supit{d},
Ian Parry\supit{f},
Laurent Pueyo\supit{g},
Emily Rice\supit{dh},
Lewis C. Roberts, Jr.\supit{a},
Jennifer Roberts\supit{a},
Michael Shao\supit{a},
Anand Sivaramakrishnan\supit{g},
Remi Soummer\supit{g},
Hong Tang\supit{a},
Tuan Truong\supit{a},
Gautam Vasisht\supit{a},
Fred Vescelus\supit{a},
J. Kent Wallace\supit{a},
Chengxing Zhai\supit{a},
Neil Zimmerman\supit{i}
\skiplinehalf
\supit{a} Jet Propulsion Laboratory, California Institute of Technology, Pasadena, CA, 91109 USA\\
\supit{b} California Institute of Technology, Pasadena, CA 91125, USA \\
\supit{c} NASA Exoplanet Science Institute, Pasadena, CA 91125, USA \\
\supit{d} American Museum of Natural History, New York, NY 10024, USA \\
\supit{e} Notre Dame University, Indiana, USA \\
\supit{f} Institute of Astronomy, Cambridge University, Cambridge, UK \\
\supit{g} Space Telescope Science Institute, Baltimore, MD 21218, USA \\
\supit{h} College of Staten Island, Staten Island, NY 10314, USA \\
\supit{i} Max Planck Institute for Astronomy, Heidelberg, Germany
}

\begin{document}
\maketitle
\pagestyle{empty}

\begin{abstract}
The Project 1640 instrument on the 200-inch Hale telescope at Palomar Observatory is a coronagraphic instrument with an integral field spectrograph at the back end, designed to find young, self-luminous planets around nearby stars.  To reach the necessary contrast for this, the PALM-3000 adaptive optics system corrects for fast atmospheric speckles, while CAL, a phase-shifting interferometer in a Mach-Zehnder configuration, measures the quasistatic components of the complex electric field in the pupil plane following the coronagraphic stop.  Two additional sensors measure and control low-order modes.  These field measurements may then be combined with a system model and data taken separately using a white-light source internal to the AO system to correct for both phase and amplitude aberrations.  Here, we discuss and demonstrate the procedure to maintain a half-plane dark hole in the image plane while the spectrograph is taking data, including initial on-sky performance.
\end{abstract}

\section{Introduction}

Indirect methods of exoplanet detection have proved wildly successful in discovering exoplanets; radial velocity and transit techniques have found hundreds of planets to date, and thousands more likely candidates.  However, these techniques have very limited sensitivity for planets at wide orbital separations, due to their long periods.  In these regions, direct imaging via coronagraphy provides a powerful tool for the detection and characterization of planets in wide orbits.

Project 1640\mycitep{Hin11, Opp12} is a high-contrast coronagraphic imaging instrument for the 200'' Hale Telescope at Palomar Observatory designed to attenuate the light from a target star in order to image dim objects in proximity, such as exoplanets and brown dwarfs, and obtain spectral information about their atmospheres.  Our ongoing 99-night survey is searching for young, warm, self-luminous planets of Jupiter mass or greater, orbiting nearby A- and F-stars, with an eventual target contrast of $10^{-7}$  at 1'' separation.

\subsection{Instrument overview}

The instrument is mounted at the Cassegrain focus of the telescope and has four major components, all necessary for successful high-contrast performance:
\begin{itemize}
\item the apodized pupil Lyot coronagraph (APLC) \mycitep{Sou11}
\item the integral field spectrograph (IFS) \mycitep{Hin08}
\item the PALM-3000 adaptive optics system (P3K) \mycitep{Bou09}
\item the Project 1640 internal wavefront sensing and control, including the calibration interferometer (CAL)
\end{itemize}

The APLC, like other coronagraphs, is designed for high level of broadband starlight suppression while leaving light largely affected from other sources (e.g. planets) at small angular separations.  In this design, an apodizer at the initial pupil reshapes the beam to maximize light in the PSF core across a wide band, which is then stopped by a hard-edged focal plane mask.  A Lyot stop, matched closely to the telescope pupil, blocks scatter induced by this mask.   

This particular coronagraph architecture was chosen due to its relative simplicity and manufacturability, as well as its compatibility with the large secondary obscuration and spiders of the Hale telescope. The hardware implementation of the APLC on Project 1640 uses a creates its apodization with microdots \mycitep{Tho08} and uses a reflective focal plane mask with a hole as the ``stop'', allowing this unwanted starlight to be picked off and used elsewhere.

The integral field spectrograph is the science camera for the instrument, providing simultaneous imaging and low-resolution spectroscopy.  For imaging, it provides a scale of 19.2 mas/pixel and gives a field of view of roughly 3.84''$\times$3.84''; for spectroscopy, the signal from the dispersed light is processed into 32 channels of $\sim25$nm width spanning Y-, J-, and H-bands ($0.99-1.78\mu$m).

The adaptive optics system uses deformable mirrors (DMs) with 3388 and 241 actuators in a woofer-tweeter configuration to correct fast atmospheric wavefront aberrations before the instrument at speeds up to 2kHz.  Sensing is done with a visible-band Shack-Hartmann with a 63x63 lenslets, with 2x2 camera pixels behind each; this light is picked off directly before entering the instrument.  This system is not specific to Project 1640, but serves other Palomar instruments (PHARO, SWIFT, TMAS) as well.

Project 1640 as has its own internal wavefront control system to correct both low- and high-order aberrations in the coronagraph itself, which vary over the course of the night due to telescope flexure and thermal variations and which P3K is not able to sense.  The primary tool is the calibration interferometer (CAL), a common-path white-light interferometer which picks off the core of the starlight as a reference to correct small, slowly-varying phase aberrations in the wavefront due to temperature changes and instrument flexure.  Two additional sensors (a quad-cell photodiode and a NIR Shack-Hartmann) are used for tip-tilt and low-order aberrations, respectively; all three will be discussed below.

Correction of wavefront aberrations in the coronagraph is essential to the full functioning of the instrument.  Sections \ref{sec:phase} and \ref{sec:amp} will discuss the estimation and correction of phase and amplitude errors in the system, respectively, and Section \ref{sec:test} will cover in-lab and on-sky tests of these wavefront control schemes.

\section{Phase correction with P1640/CAL} \label{sec:phase}

Prior to performing amplitude correction, we first perform low- and high-order phase corrections.  Three sensors are used for this: one for tip-tilt, one for low-order errors (other than tip-tilt), and one for high-order corrections.

For tip-tilt correction, the Y-band and J-band portions of the beam are picked off at a dichroic beam splitter and repointed onto the sensor, which consists of four Hamamatsu InGaAs photodiodes mounted as a quad cell in a single package.  To initialize the sensor, the beam is centered on the focal plane mask using a combination of CAL camera images and IFS cubes; an actuated mirror is then used to center the beam on the quad cell.  A PID control loop, written in LabView, then uses a fast steering mirror near the apodizer to keep the signal in the four cells balanced.

\begin{figure}
\begin{center}
\includegraphics[width=3.0in]{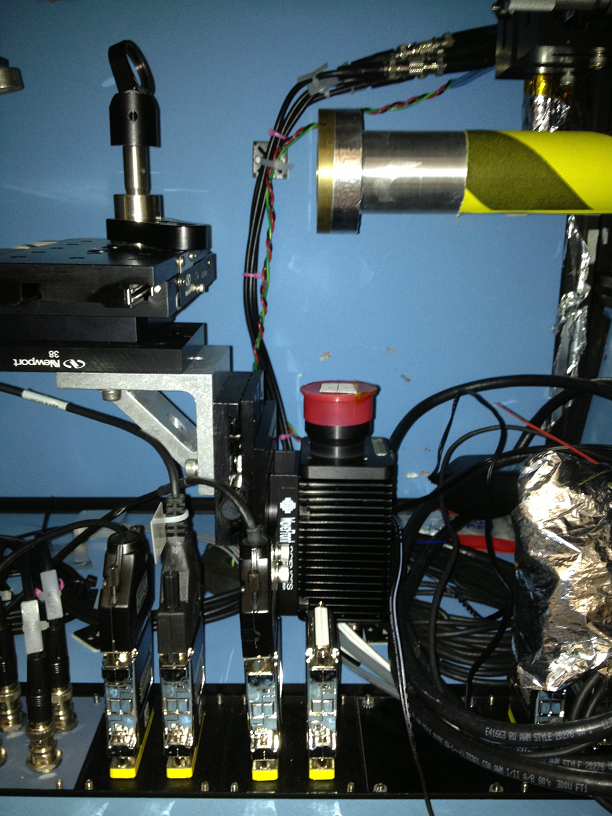}
\caption{The low-order Shack-Hartmann sensor, with red lens cap present at center.}  \label{lowfs}
\end{center}
\end{figure}

Low-order correction is done with a new sensor, a self-contained near-IR Shack-Hartmann made by Wavefront Sciences, with 20x16 lenslets.  Given throughput considerations and space constraints within the instrument, this sensor could not be placed in the beam path; rather, we use the actuators on the focal plane mask to tilt the beam at an image plane, shifting the beam onto a small 45$^{\circ}$ mirror which sends the post-coronagraphic science beam out of plane to the Shack-Hartmann in a conjugate pupil.  The beam is approximately 8x8 lenslets, and the wavefront is measured using the proprietary control software for the camera.  The wavefront phase is then shifted, rotated and interpolated to match the DM sampling, and a DM setting is applied to directly conjugate the measured phase.  An image of the vertically-mounted camera is shown in \myfig{lowfs}.  Residuals are generally found to 20-30nm peak-to-valley of low-order aberration.

High-order correction is done with the main wavefront correction instrument, the CAL interferometer.  CAL is a common-path white-light interferometer arranged in a Mach-Zehnder configuration; it interferes a filtered beam from the core of the star’s point spread function (the reference beam) with $20\%$ of the post-coronagraphic wavefront, picked off before the Lyot stop with a beamsplitter (the science beam).  A filter at the final camera limits the measurement to H-band.

The path length in the reference beam is modulated in four $\lambda/4$ steps. The CAL camera is conjugate to the pupil; we can measure phase at speeds from 0.2Hz to 5Hz and send commands to conjugate it once per minute.  Four interfered intensity images (A, B, C, D) directly give the amplitude and phase of the science beam in the pupil as:
\begin{eqnarray}
\mathrm{amplitude} =& \sqrt{(A-C)^2 + (B-D)^2} \\
\mathrm{phase} =& \arctan{\frac{A-C}{B-D}}
\end{eqnarray}
A more detailed description of the wavefront extraction, including calibration and DM registration, is given in \mycitet{Zha12}.

Phase conjugation is in principle simple: the DMs are commanded to move to $-1/2$ of the residual phase value.  These commands are sent to P3K as 66x66 actuator DM commands, and ordered to be interpreted as centroid offsets in the P3K wavefront sensor when the adaptive optics loops are closed, and as direct DM pokes otherwise.  After an initial calibration sequence, the correction procedure is entirely automated, requiring only a number of iterations to be entered. 

Our best on-sky performance as of July 2013 is 5nm phase residuals on H=5-6 stars; however, we still remain limited by $\sim7\%$ residual amplitude error. $3\%$ is random, while the remainder ($\sim4\%$) is residual diffraction. 

\section{Amplitude correction with P1640/CAL} \label{sec:amp}

Amplitude cannot be conjugated directly as phase can to bring down the quasistatic speckle floor across the controllable region of the image plane; rather, it must be corrected across a part of that region, most often a half-plane. Two schemes have been implemented to minimize the intensity within a half-plane region of the image plane for Project 1640: Energy Minimization \mycitep{Bor06} and Electric Field Conjugation (EFC; see \mycitet{Giv09}).  

For energy minimization, we find a correction to the image-plane field $E_{aberrated}$ by considering only the field in a half-plane region, and replacing the field in the other half with the Hermitian conjugate.  The inverse Fourier transform of this gives a real-valued correction, which is applied iteratively.  While this approach can take many iterations to converge, it requires no knowledge or models of the system. This has been tested successfully in the laboratory, and further discussions may be found in \mycitet{Zha12}.

For EFC, assuming a set of actuator heights $a$, we write $E_{corrected}$ = $E_{aberrated} + Ga$, with $G$ a matrix containing effect of each actuator poke on the image-plane electric field.  In principle, $G$ can be determined by either by models or by direct pokes in advance, although we have chosen to use models, based on the expected time to complete a full set of pokes.  The naive algorithm then attempts to find an $a$ which minimizes the 2-norm $||E_{corrected}||_2$ by choosing 
\begin{equation}
a = -\left[
\begin{array}{c}
\mathfrak{R}(G) \\
\mathfrak{I}(G) \\
\end{array}\right]^{\dagger} \left[
\begin{array}{c}
\mathfrak{R}(E_{aberrated}) \\
\mathfrak{I}(E_{aberrated}) \\
\end{array}\right]
\end{equation}
with $^{\dagger}$ denoting a left pseudoinverse, and the $\mathfrak{R}$ and $\mathfrak{I}$ representing real and imaginary parts.  A standard extension adds Tikhonov regularization with an adjustable parameter $\mu$ to limit the effect of model errors.  Further modifications can allow for broadband corrections given field estimates at multiple wavelengths, but these are not implemented currently, given that the CAL camera does not have the capability to image at multiple wavelengths.

This scheme can converge very quickly and produce quite deep nulls, but requires a good model of the system to work effectively. (For either algorithm, a minimal amount of modeling is required, as CAL measures pupil plane fields and corrects image plane regions.)  System modeling poses a particular problem for Project 1640, as the Lyot stop is not common-path with the CAL camera; it must be centered as best possible using the IFS prior to use.  

\begin{figure} 
\begin{center}
\includegraphics[width=0.7\columnwidth]{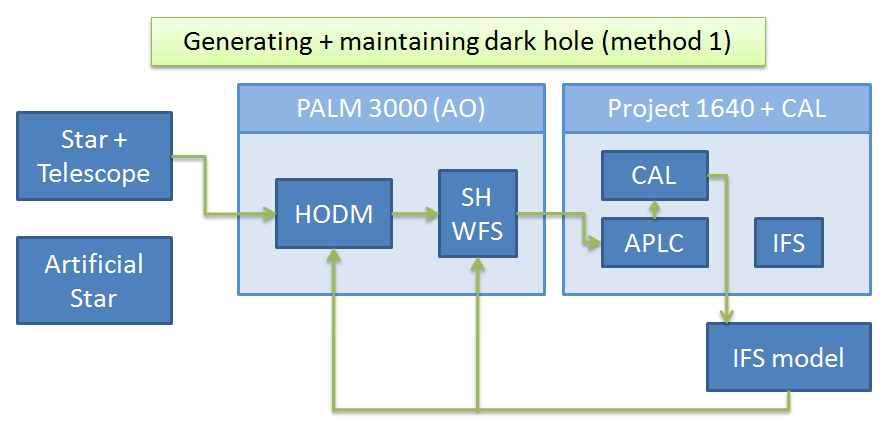}
\caption{A block diagram of the existing phase and amplitude control loops for Project 1640.} \label{method1}
\end{center}
\end{figure}

The current control loop is shown in \myfig{method1}; note that this scheme does not require feedback from the science camera to perform either phase or amplitude estimation and control.  A common modification is to get initial phase and/or amplitude corrections with the artificial star in the P3K system prior to switching to sky, and then touch up the dark hole from there; this generally saves time under actual observing conditions.  As with phase conjugation, running EFC is entirely automated, assuming the initial calibration and G-matrix are fed in.

\section{Testing} \label{sec:test}

\begin{figure}
\begin{center}
\subfigure{
\includegraphics[width=3.25in]{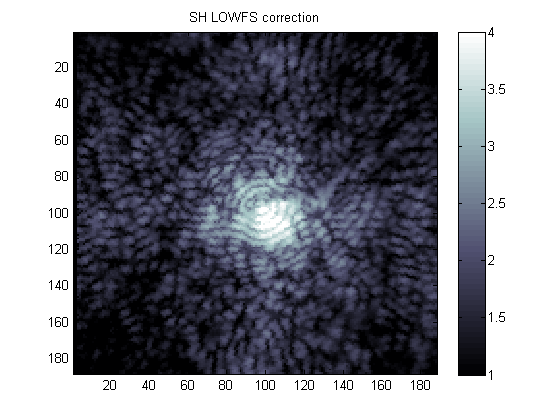}
\includegraphics[width=3.25in]{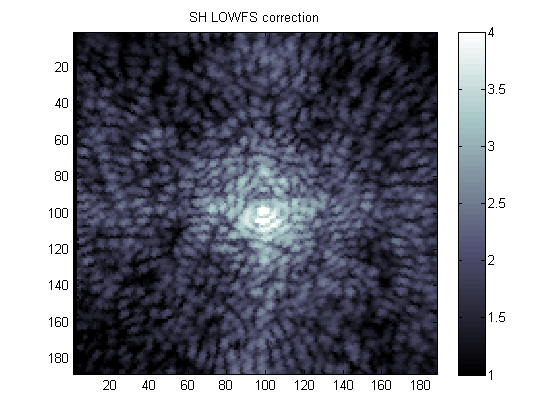}
} \\
\subfigure{
\includegraphics[width=3.25in]{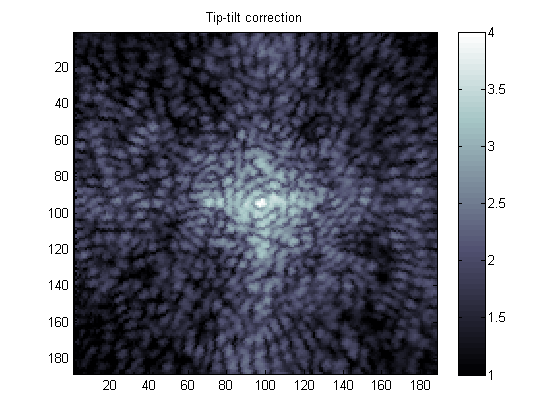}
\includegraphics[width=3.25in]{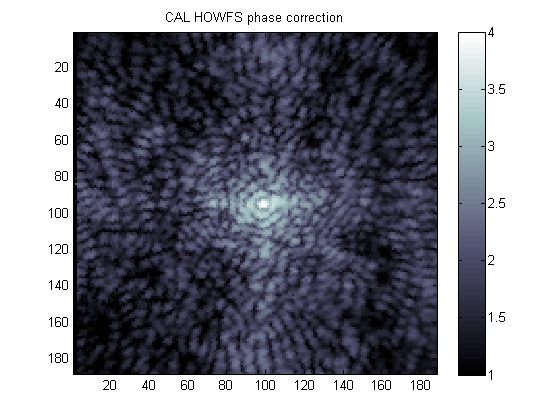}
} \\
\subfigure{
\includegraphics[width=3.25in]{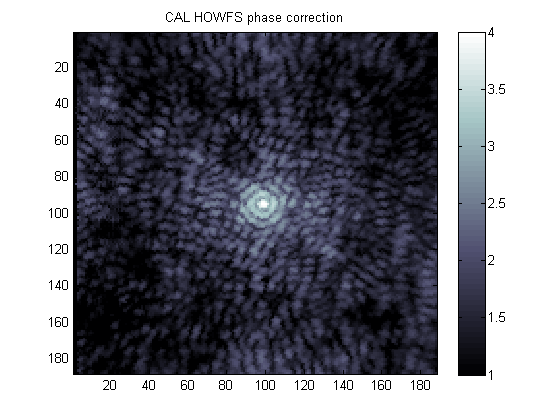}
\includegraphics[width=3.25in]{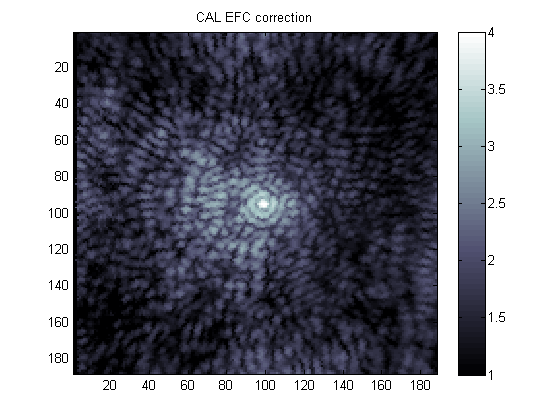}
}
\caption{These six images, taken as part of a full correction sequence, show the various stages of phase and amplitude correction on the white-light source.  All are log-scale, with identical normalization and color scale.  \emph{Top left.} Prior to the start of low-order correction \emph{Top right.} After the completion of low-order correction. \emph{Middle left.} After the locking of the tip-tilt loop.  \emph{Middle right.}  After the first high-order phase conjugation step.  \emph{Bottom left.} After the completion of high-order phase conjugation.  \emph{Bottom right.} After two iterations of EFC on the right half-plane.
}  \label{sixstep}
\end{center}
\end{figure}

The majority of the testing has been performed using the P3K white light source, which is our standard light source while obtaining initial quasistatic phase and amplitude corrections.  For example, \myfig{sixstep} shows a sequence of IFS cube slices, taken during a cloudy night in May 2013 observing run, illustrating the different steps of phase and amplitude correction as implemented.

\begin{figure}
\begin{center}
\includegraphics[width=3.25in]{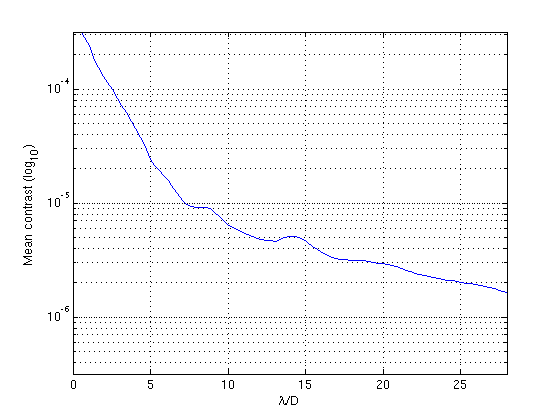}
\caption{Photometrically-calibrated mean contrast on the P3K white-light source, showing the current level of residual H-band speckle in the IFS following phase correction, as a function of radius.}  \label{labp4}
\end{center}
\end{figure}

The power of this source may be changed from the P3K control GUI, which, combined with control over the IFS exposure time, allows us to photometrically calibrate the dark hole depths as measured directly from the IFS.  Using a sequence of images with the PSF core both aligned and misaligned from the FPM center, we bootstrapped from an unsaturated core to an occulted long-exposure image during an engineering run in May 2013; the mean quasistatic speckle level at 1623nm, after phase correction, is shown in \myfig{labp4} as a function of angular separation. 

\begin{figure}
\begin{center}
\subfigure{
\includegraphics[width=3.25in]{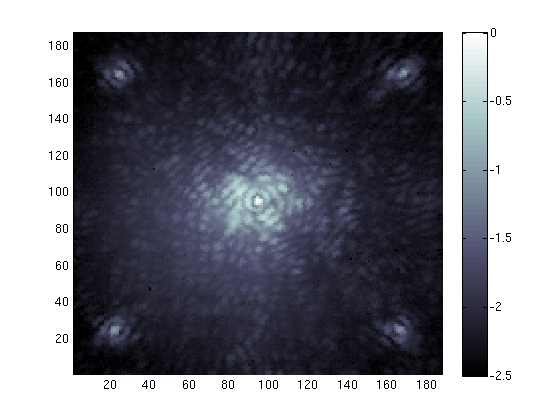}
\includegraphics[width=3.25in]{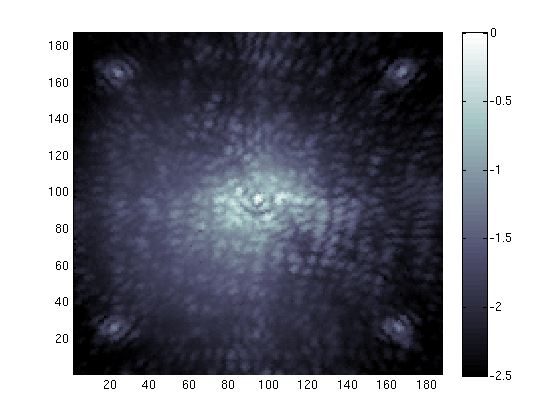}
}
\caption{\emph{Left.} A slice of a phase-only image taken on a P1640 science target under clear conditions. \emph{Right.}  A slice of a phase-and-amplitude image taken on Vega under lightly-cloudy conditions.  Both have the same quasistatic phase correction, with an additional amplitude correction being applied on Vega for the upper half-plane.  Both images are log-scale and normalized to their respective PSF peaks; units are in pixels.} \label{vegaAndSci}
\end{center}
\end{figure}

During the night of 7/21/13, we tested the application of a half-plane dark hole successfully on the sky. \myfig{vegaAndSci} shows two cube slices taken from these images.  Observations were initially taken on the star Vega during lightly-cloudy conditions, using amplitude and phase correction settings precomputed using the white-light source for a half-plane dark hole in the upper half of the image.  However, a sudden clearing in the clouds led us to switch to a nearby science target and reload a CAL correction containing the phase corrections applied on Vega, but no amplitude corrections.   (Subsequent heavy clouds and high relative humidity caused no further data to be taken on that observing run.)  Thus, while we have before-and-after images with and without amplitude correction, the two cannot be directly compared photometrically, as they were taken on different stars.  Nonetheless, we can look at the differences between upper and lower halves to examine on-sky performance.

\begin{figure} 
\begin{center}
\subfigure{
\includegraphics[width=3.25in]{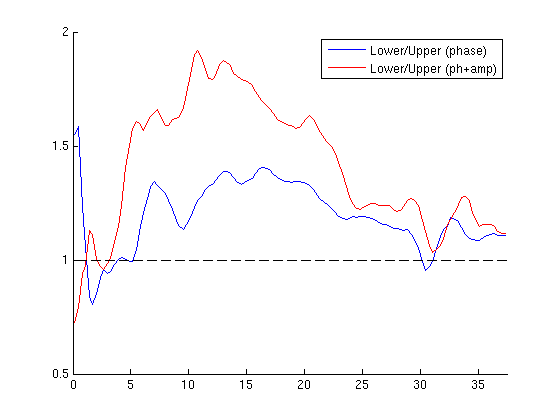}
\includegraphics[width=3.25in]{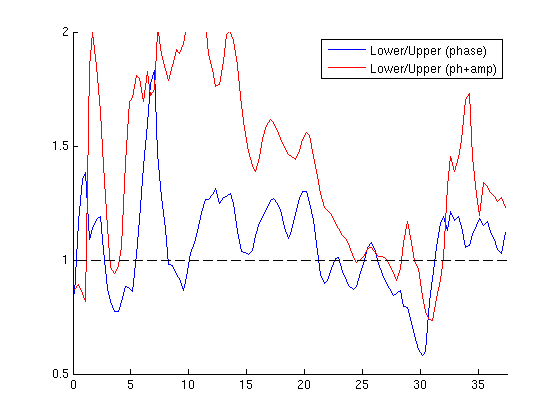}
}
\caption{\emph{Left.} Ratios of means between the upper and lower halves of the image plane, for the phase-only and phase-and-amplitude cubes. \emph{Right.} Ratios of standard deviations between the upper and lower halves of the image plane, for the phase-only and phase-and-amplitude cubes. X-axis units are in $\lambda/D$.} \label{stds}
\end{center}
\end{figure}

In ground-based imaging, the smooth halo from atmospheric residuals degrades the applicability of mean contrast when on-sky; 5-$\sigma$ contrast serves as better measure of detectability.  (5-$\sigma$ contrast at a given angular separation is estimated from $5\times$ standard deviation of all points in a $1\lambda/D$-wide annulus centered at the desired separation.) \myfig{stds} shows the ratio between both mean and 5-$\sigma$ contrasts in the upper- and lower-half annuli as function of angular separation, both as ratios of means and standard deviations.  The ratio for the phase-only data is shown as a reference, as well.  The primary improvement can be seen at mid-spatial frequencies in both cases; the limit, seen when iterating repeatedly with EFC, is the small final non-common-path between CAL and the IFS, particularly the Lyot stop.  

\section{Future directions}

We expect subsequent work on amplitude control to follow two paths.  First, we will continue work on the existing algorithm.  Currently, $G$ matrices are precalculated and loaded based on the desired dark hole geometry and regularization; they are not using the additional information given by the low- and high-order phase corrections.  Some additional code development, both in the model (for speed) and the low- and high-order phase correctors (to maintain a consistent estimate of total applied DM settings) will be required for this.  We also expect to use this algorithm to target specific small regions of the image plane with suspected targets, with the hope of improving the signal-to-noise ratio of the extracted spectrum.

\begin{figure} 
\begin{center}
\includegraphics[width=0.7\columnwidth]{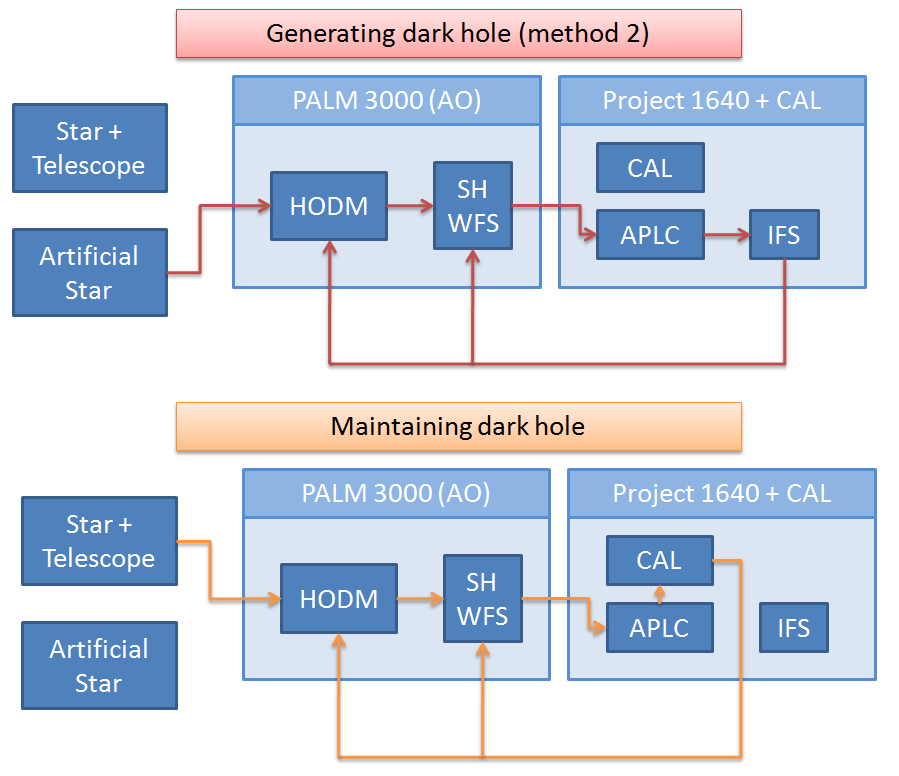}
\caption{A block diagram of a planned phase and amplitude control scheme for Project 1640, permitting broadband control, no non-common-path errors, and maintenance during observation.}\label{method2}
\end{center}
\end{figure}

Second, we will investigate an alternate control-loop architecture, shown schematically in \myfig{method2}, which we hope will eliminate our difficulties with non-common-path error.  In this scheme, a half-plane dark hole is first created with the white-light source and the IFS cubes, using one (or more) of the amplitude control schemes already implemented for CAL.  Using the IFS not only eliminates the challenge of correctly placing the Lyot stop in the model with no feedback, it permits the dark hole to be made in Y- and J-bands, in addition to H-band.  Using the white-light source obviates the difficulty of performing these measurements on sky, due both to low photon counts and the presence of atmospheric turbulence residuals.

The second half of this scheme is to measure the wavefront $W_{t}$ associated with this new dark hole using CAL, and switch to on-sky observation.  Once there, we  use CAL to minimize the difference between the measured wavefront and $W_{t}$, locking the dark hole in place while observing without requiring any wavefront estimation to be done in the IFS.

This architecture has been recently enabled by the continued maturation of the cube extraction pipeline, which is now becoming fast enough to permit the creation of the dark hole on IFS during the night. Significant work is still required to automate the calibration and extraction of wavefronts from the IFS, though, as well as linking the currently-distinct IFS and CAL software.

Overall, however, electric field estimates from the CAL wavefront sensor may be used now to generate dark holes over full-plane regions (with phase correction) and half-plane regions (with both amplitude and phase correction). Phase correction is already used for on-sky correction; amplitude correction has been tested on sky, and we expect to bring it into regular use in upcoming observation runs.

\section*{Acknowledgments}

This work was performed in part at the Jet Propulsion Laboratory, California Institute of Technology, under contract to the National Aeronautics and Space Administration.

\bibliography{refs}
\bibliographystyle{spiebib}

\end{document}